\documentclass[apj]{emulateapj}

\usepackage{color}

\begin{document}

\voffset -1cm

\title{Can life survive Gamma-Ray Bursts in {the} high-redshift universe?}

\author{Ye Li, Bing Zhang}

\affil{Department of Physics and Astronomy, University of Nevada, Las Vegas, NV 89154, USA \\
liye@physics.unlv.edu; zhang@physics.unlv.edu}

\begin{abstract}

Nearby Gamma-Ray Bursts (GRBs) have been proposed as
{a possible} cause of mass extinctions on Earth.
Due to the higher event rate of GRBs at higher redshifts, it has been speculated that
life as we know {it} may not survive above a certain redshift (e.g. $z>0.5$).
We examine the duty cycle of lethal (life-threatening) GRBs in the solar neighborhood, in
the Sloan Digital Sky Survey (SDSS) galaxies and 
GRB host galaxies, with the dependence of the long GRB rate on star formation and
metallicity properly taken into account.
We find that the number of lethal GRBs attacking Earth within the past
500 Myr ($\sim$ epoch of the Ordovician mass extinction) is $0.93$.  
The number of lethal GRBs {hitting a certain planet}
increases with redshift, {thanks to the increasing
star formation rate and decreasing metallicity in high-$z$ galaxies.}
Taking 1 per 500 Myr as a conservative duty cycle for life to survive, as evidenced by our existence,
we find that there are still a good fraction of SDSS galaxies beyond $z=0.5$
{where the GRB rate at half-mass radius is lower than this value.}
We derive the fraction of such benign galaxies as a function of redshift through Monte Carlo simulations,
and find that the fraction is $\sim 50\%$ at $z\sim 1.5$ and $\sim 10\%$ even at $z \sim 3$.
{The mass distribution of benign galaxies is dominated by Milky-Way-like ones,
thanks to their commonness, relatively large mass,
and low star formation rate. 
GRB host galaxies are among the most dangerous ones.}

\end{abstract}


\section{Introduction}
Nearby high-energy transient sources have been considered as a possible cause of
mass extinction events on Earth as well as a potential threat {to} life in the future
{\citep{1974Sci...184.1079R, 1995ApJ...444L..53T, 1998PhRvL..80.5813D,
2003ApJ...585.1169G, 2011AsBio..11..343M, 2012MNRAS.423.1234S}.}
Among them,
gamma-ray bursts (GRBs), the most violent explosions in the universe, have been
regarded as {one of} the most lethal high-energy transients to life
\citep{2002ApJ...566..723S, 2005ApJ...622L.153T, 2005ApJ...634..509T}.
With a 2-D simulation on how Earth{'s} atmosphere responds to an intense $\gamma$-ray
flux, \cite{2005ApJ...622L.153T} investigated the effect of a nearby (2 kpc away) GRB
{to} life on Earth. A typical GRB at this distance has 
a $\gamma$-ray fluence of $10^8~ {\rm erg~ cm^{-2}} ~(= 100 ~{\rm kJ~m^{-2}})$,
which would cause severe damage to life on Earth. 
According to \cite{2005ApJ...622L.153T},
such a GRB would lead to a rapid increase of nitrogen compounds 
(e.g. NO and NO$_2$) in the atmosphere 
causing an on-average 35\% ozone depletion in the stratosphere, 
{increasing for years the solar
UVB radiation flux at Earth.
The resulting DNA damage, up to 16 times the normal level,
is lethal to many organisms such as plankton, the base of the food chain.}
It may lead to
{extinction of creatures in higher trophic levels of the food} chain due to starvation. 
Furthermore, the opacity of NO$_2$ in {Earth's} atmosphere would result in a decreased surface temperature, which {is speculated to 
be a cause of a long-lasting ice age (\citealt{2005ApJ...622L.153T}, but see
\citealt{2015AsBio..15..207T}).}
The mass extinction in {the} late Ordovician ($\sim$ 447 Myr ago)
could be due to the impact of
a nearby GRB \citep{2004IJAsB...3...55M, 2006AREPS..34..127B, 2008arXiv0809.0899M}.

{Long-duration GRBs are more dangerous than
the short-duration ones \citep{2014PhRvL.113w1102P}}. Since long GRBs (LGRBs hereafter)
are related to
deaths of massive stars and therefore track the star formation history of the universe
\citep{2006ARA&A..44..507W, 2015PhR...561....1K},
it is naturally expected that GRBs become 
more lethal at higher redshifts where the star formation rate is higher. 
Indeed,  \cite{2014PhRvL.113w1102P} speculated that life as we know {it} cannot
survive at $z>0.5$ due to the frequent bombardment of GRBs at any location
in a galaxy.

One caveat in drawing such a conclusion is related to the duty cycle of lethal GRBs {preventing development of advanced life}. The time scale to re-develop advanced life is
not well studied. 
Even though {the} Ordovician mass extinction happened $\sim$ 447 Myr ago,
{more recent mass extinction events (e.g. the Cretaceous-Paleogene extinction $\sim$ 65 Myr ago that killed the dinosaurs) did not prevent re-emergence of advanced life forms (humans).}
{Therefore it may be possible for life to tolerate a shorter (say, 50 Myr) lethal GRB duty cycle.} {Regardless} of the biological details of how a GRB may kill life and how advanced life
forms re-emerge, our existence suggests that advanced civilizations can develop  
if lethal GRBs have a duty cycle comparable to the one inferred for the solar neighborhood
in the Milky Way Galaxy. 

In this paper, we quantify the lethal GRB duty cycle in the solar neighborhood,
and apply it as a conservative life survival condition to study the ``habitability''
of various observed galaxies (e.g. Sloan Digital Sky Survey [SDSS] galaxies and
the GRB host galaxies). The aim is to address whether life can survive GRBs in the
high-redshift universe.

\section{Methodology}

Following
\cite{2005ApJ...622L.153T}, 
we adopt $F_{\rm c}=10^8\ {\rm erg\ cm^{-2}}$ as a critical fluence {defining} a lethal GRB.
Since GRBs trace star formation and most star formation 
happens in late type galaxies,
we consider our test galaxies as disk galaxies 
with an exponential stellar mass column density
$\Sigma_*e^{-r/r_{d}}$, where $r_{d}$ is the scale length.
To the first-order approximation, one may ignore variation of 
specific star formation rate (sSFR) and metallicity within the galaxy.
A GRB at $r_0$ from the galactic center defines a pair of cones where
life is damaged. These cones have a radius
$R(L, F_c)=\sqrt{L \Delta t/(4\pi F_{\rm c})}$ and a solid angle $\Omega$,
where 
$L$ is the peak isotropic luminosity of the GRB, 
and $\Delta t$ is the rest-frame duration of the GRB, which is typically $\sim 10$ s.
For a GRB at $r_0$, the fraction of mass in the galaxy where life is 
damaged can be expressed as
\begin{eqnarray}
p\left(L, r_0\right) & = & p[R(L,F_{\rm c}), r_0] = \frac{\int \Sigma_* e^{-r/r_d} dA}{M_*} \nonumber \\
&=& \int^{^{r_0+R}}_{_{r_0-R}} dx 
\int^{^{\sqrt{R^2-(x-r_0)^2}}}_{_{-\sqrt{R^2-(x-r_0)^2}}}
\Sigma_* e^{-\frac{\sqrt{x^2+y^2}}{r_{d}}} dy \nonumber \\
 & \times & f_{b} ~M_*^{-1}, 
\label{eq:p}
\end{eqnarray}
where $f_{b}=\frac{\Omega}{4\pi}$ is the beaming correction factor.
The fraction $p(L,r_0)$ is also the probability for a random
GRB in the galaxy (including both those beaming towards and those beaming away from the planet)
to kill life on a planet at $r_0$ from the galactic center.

The high-luminosity {LGRB} luminosity function (LF), $\phi(L/L^*)$ 
is characterized by a broken power law with a break luminosity $L^*$
\citep{2007ApJ...662.1111L,2010MNRAS.406.1944W,Sun15}, e.g.
\begin{equation}
\phi \left( \frac{L}{L^*} \right) \propto
\left[ \left( \frac{L}{L^*} \right)^{\alpha_1}+
\left( \frac{L}{L^*} \right)^{\alpha_2} \right]^{-1}. 
\end{equation}
The number of lethal GRBs that would attack the planet at $r_0$
within a time duration $T$ is  
\begin{eqnarray}
\!\!\!\!\!N & = & \int^{L_{\rm max}}_{L_{\rm min}} \!f_{b}^{-1} \dot\rho_0
\phi (L) p(L,r_0) d L\cdot T V(M_*) f_{\rm sSFR} f_{\rm  Fe}, 
\label{eq:N}
\end{eqnarray}
where the range of integration is $10^{49} - 10^{55} \rm erg\ s^{-1}$,
the observed range of LGRBs, and
$\dot \rho_0$ is the GRB event rate density above a specific luminosity 
in the local universe. 
Here the factor $f_b^{-1}$ makes the correction from the observed GRBs (defined
by the observed luminosity function) to the total GRBs (including those not beaming
towards us). Notice that it cancels out with the $f_b$ factor in Eq.(\ref{eq:p}),
so that the result does not depend on the poorly constrained parameter $f_b$.

Various studies reach a generally consistent conclusion, but with somewhat different
parameters. Here we adopt the latest analysis by \cite{Sun15} using the largest 
long GRB sample. Since there is an evolution of the GRB luminosity function, we
use the luminosity function and event rate density 
derived from the nearby ($z<1$) sample with the following parameters
(\citealt{Sun15}):
$\alpha_1 = 1.57$, $\alpha_2 = 1.8$, $L^*=4\times 10^{51} {\rm\ erg\ s^{-1}}$,
and $\dot \rho(>10^{50}~{\rm erg~s^{-1}}) = 1.6~{\rm Gpc^{-3} yr^{-1}}$.
The {occupied cosmological volume} of a galaxy with mass $M_*$ is estimated as 
$V(M_*)={M_*}/{\rho_*(z)}$, where
$\rho_*(z)$ is the average stellar density obtained by integrating 
the stellar mass function \citep{2015MNRAS.447....2M} from $10^7 - 10^{13} M_{\sun}$ 
for different redshifts, which is fit as $\rho_*(z)=10^{17.46-0.39z}\ \rm M_{\sun}\ Gpc^{-3}$
(see also 
\cite{2013ApJ...777...18M} and papers therein).
A typical time scale $T$ = 500 Myr is used to {match approximately} the epoch of the
Ordovician Mass Extinction {of} $\sim$ (447 -- 443) Myr ago
\citep{2000Geo....28..967S, 2003GSAB..115...89B}.

The parameter $\dot\rho_0$ denotes the average GRB event rate density in the local
universe ($z \sim 0$). We know that the {LGRB} rate depends on star formation
rate and metallicity. It is then relevant to introduce two correction factors
for the specific galaxy values with respect to the local average values.
The specific SFR correction factor is introduced as 
\begin{equation}
f_{\rm sSFR}=\frac{\rm sSFR}{\rm sSFR_0},
\end{equation}
where the local specific star formation rate is sSFR$_0= 0.1\ {\rm Gyr^{-1}}$
\citep{2011MNRAS.417.2737W}.
The metallicity correction factor is defined as 
\begin{equation}
f_{\rm Fe}=\frac{P({\rm [Fe/H]})}{P_0({\rm [Fe/H]_0})}.
\end{equation}
Here $P({\rm [Fe/H]})$ is the fraction of stars with metallicity 
poorer than [Fe/H] $= -0.43$ (equivalent to 0.4 $Z_{\sun}$)
\citep{1994A&AS..106..275B, 2011MNRAS.417.3025V},  
assuming that the metallicity of the galaxy is a Gaussian distribution
with the detected metallicity as the medium value 
and the standard deviation $\sigma_{\rm Fe}=0.22$, similar to the Milky Way
\citep{2011A&A...530A.138C}.
The function $P({\rm [Fe/H]_0})$ is a similar fraction, but
with median value being the mean metallicity of the local universe
${\rm [Fe/H]_0}=-0.006$ 
\citep{2008MNRAS.383.1439G, 2014ARA&A..52..415M}.

The short GRB (hereafter SGRB) impact rate can be estimated with a similar method. 
The typical duration of a SGRB $\Delta t$ is $\sim$ 0.5 s.
The luminosity function depends on the model of {the} merger delay time distribution,
but can be generally fitted as a single power law \citep{Sun15}. We use the
best-fit luminosity function parameters for a Gaussian delay time distribution, 
i.e. $\phi(L) \propto L^{-1.7}$ with $\dot\rho_0 (>10^{50}~{\rm erg~s^{-1}})
=1.3~{\rm Gpc^{-3}~yr^{-1}}$.
The range of integration is $10^{49}-2\times10^{54} \rm{\ erg \ s^{-1}}$ as observed.
Since the SGRBs are {likely} due to mergers of two neutron stars or
a neutron star--black hole system, 
the probability of a short GRB is no longer directly related to the fraction
of the mass.
Rather, we derive the probability by introducing an offset distribution of the
afterglow with respect to the host galaxies. Using the observational data
\citep{2010ApJ...708....9F, 2013ApJ...776...18F},
we find that the normalized (in {units of the host galaxy scale length}) offset 
distribution 
can be {approximated} as a Gaussian function in logarithmic space, 
i.e., G(${\rm log_{10}}(r/r_{d})$, $\mu$=0.32, $\sigma$=0.57).
So the probability of a SGRB {located in an area} $dA$ is 
\begin{eqnarray}
p(L, r_0) & = & \frac{\int G(\log_{10}(r/r_d)) dA}{(\ln 10)2\pi r^2} \nonumber \\
& \times & \frac{1}{\int G(\log_{10}(r/r_d))d(\log_{10}(r/r_d))},
\end{eqnarray}
where the integration in the numerator is over the damaging region around $r_0$
and the integration in the denominator is over the entire galaxy.
Also there is no direct connection between the SGRB rate and sSFR or metallicity,
so that $f_{\rm sSFR, SGRB} =  1$ and $f_{\rm Fe, SGRB} = 1$ are adopted.

\section{Lethal GRB RATE IN MILKY WAY}

\begin{figure}[!htb]
\centering
\includegraphics[width=1.0\columnwidth]{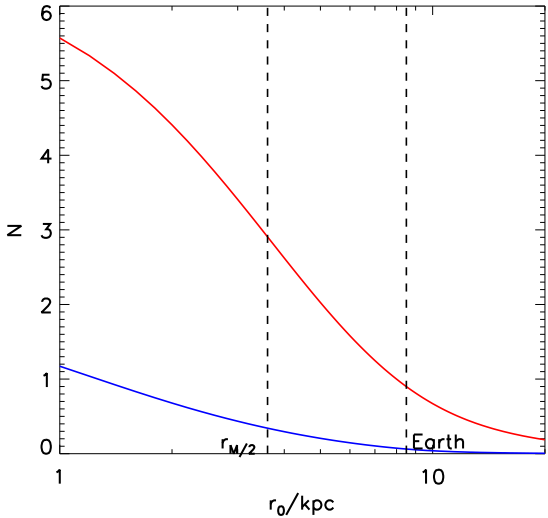}
\caption{The number of lethal LGRBs (red solid line) and SGRBs (blue solid line)
within 500 Myr as a function of distance $r_0$ from the Galactic Center.
Two vertical black dashed lines indicate the half-mass radius
($r_{\rm M/2}$) and the Earth position, respectively.
}
\label{fig:MW}
\end{figure}

We first apply our methodology to {the} Milky Way (MW).
The stellar mass of {the} MW is 
$M_*=6.08 \pm 1.14 \times 10^{10}\ \rm M_{\sun}$
\citep{2011MNRAS.414.2446M}, 
so that the {occupied cosmological volume} is $V=2.1\times 10^{-7}\ \rm Gpc^{-3}$.
With sSFR = $2.71 \pm 0.59 \times 10^{-2}\ \rm Gyr^{-1}$
\citep{2014arXiv1407.1078L},
which is relatively small in the local universe, 
we get the sSFR correction factor $f_{\rm sSFR}=0.27$.
The metallicity distribution function of {the} MW is 
a Gaussian with the mean value ${\rm [Fe/H]}=-0.06$ and $\sigma_{\rm Fe}=0.22$
\citep{2011A&A...530A.138C}.
It is relatively metal poor {compared to} the average value in the local universe, and 
the metallicity correction factor is $f_{\rm Fe}=1.7$.
The half mass radius of {the} MW is $r_{\rm d} = 2.15 \pm 0.14\ {\rm kpc}$
\citep{2013ApJ...779..115B}, and our Earth is located {at}
$r_0=8.33 \pm 0.35\ \rm kpc$ away from the Galactic center
\citep{1993ARA&A..31..345R, 2003ApJ...597L.121E, 2009ApJ...692.1075G}.

The red line in Fig.\ref{fig:MW} denotes the number of lethal LGRBs  
for planets at {different radii} $r_0$ in {the} Milky Way within 500 Myr.
At {Earth's location},  the number of lethal LGRBs is 0.93 within 500 Myr.
This is consistent with the hypothesis connecting the Ordovician mass 
extinction {with GRB activity}. The number is larger than 1 for $r_0 < 8$ kpc.
At the half mass radius, the number of lethal GRBs is $N_{\rm M/2}= 2.91$.
The number increases to even larger values at smaller radii, suggesting
that the regions close to the Galactic center {are} less habitable. 
Our existence suggests that $N \sim 1$ per 500 Myr can be regarded as a
{\em conservative} criterion for the survival of advanced life forms.

The blue line in Fig.\ref{fig:MW} shows the number of lethal SGRBs as a 
function of $r_0$ per 500 Myr. They are much rarer than LGRBs, suggesting that the
LGRBs are the dominant species {limiting} life in the universe
\citep[see also][]{2014PhRvL.113w1102P}.
In the rest of the paper, we ignore the contributions of SGRBs.

\section{Lethal GRB rate in other galaxies}

Galaxies at high redshifts have higher sSFR
\citep{2011MNRAS.417.2737W, 2014arXiv1410.4875I}
and lower metallicity
\citep{2008A&A...488..463M, 2010MNRAS.408.2115M} on average.
These result in higher $f_{\rm Fe}$ and $f_{\rm sSFR}$ correction factors,
so that the number of lethal GRBs is expected to increase with redshift.
In this section, we investigate the lethal GRB rate for other galaxies
making use of the observational data directly.
We employ the SDSS DR8 and DR12 galaxy samples
\citep{2011ApJS..193...29A, 2013AJ....146...32S, 2015arXiv150100963A} 
as the starting point, and extrapolate the sample to even higher redshifts
with a Monte Carlo simulation, aiming at quantifying the redshift-dependence
of the fraction of benign galaxies where life as we know {it} can survive GRBs.
GRB host galaxies are also studied for comparison.
For all the galaxies, instead of studying the $r_0$-dependent lethal GRB rate,
we only investigate the rate at the half-stellar-mass radius 
$N_{\rm M/2}$ as a {representation} of the ``habitability'' of the galaxy.

\subsection{SDSS DR8 and SDSS/BOSS DR12 Samples}

\begin{figure*}[!htb]
\centering
\includegraphics[width=1.0\columnwidth]{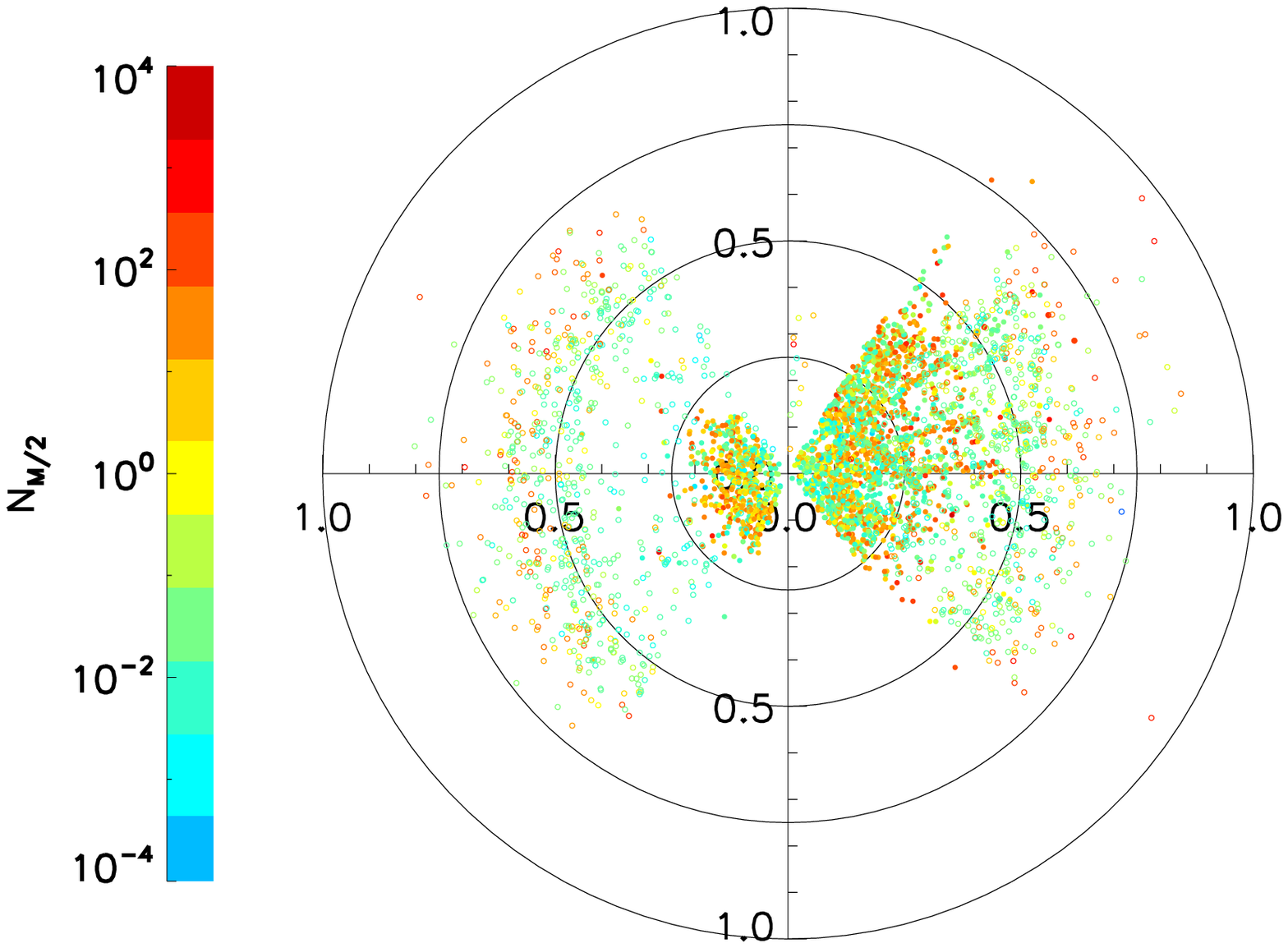}
\includegraphics[width=1.0\columnwidth]{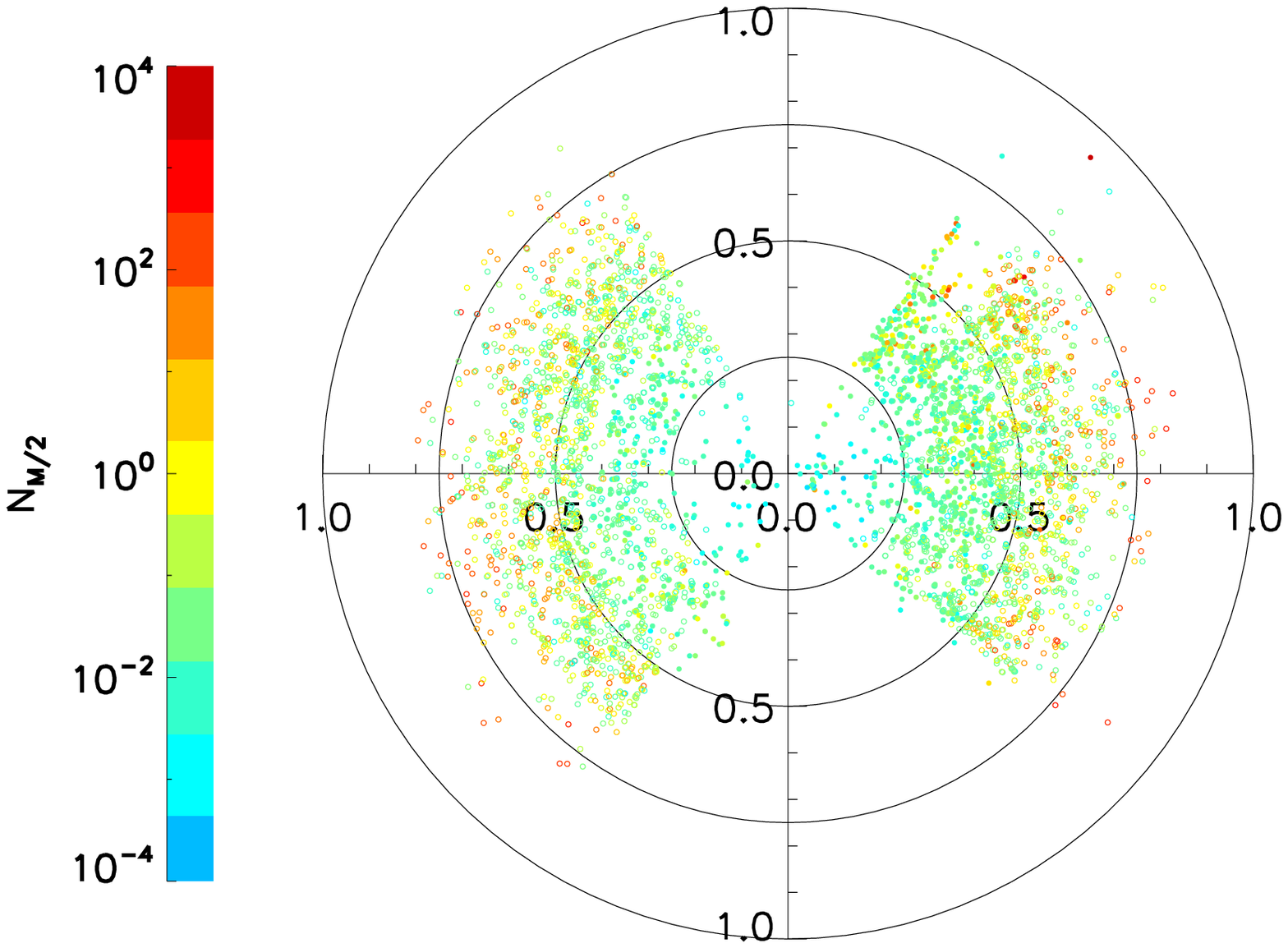}
\caption{3-D maps of SDSS DR8 (filled circles) and DR12 (open circles) galaxies
with Dec in the range [$0^{\circ}$, $0.5^{\circ}$].
Colors are encoded with the number of lethal GRBs within 500 Myr
at the half mass radius $N_{\rm M/2}$
for each galaxy. The larger the $N_{\rm M/2}$, the more dangerous the galaxy.
Left Panel: $10^{10}\ M_{\sun} < M_* <10^{11}\ M_{\sun}$;
Right Panel: $10^{11}\ M_{\sun} < M_* < 10^{12}\ M_{\sun}$.
}
\label{fig:lifemap}
\end{figure*}

\begin{figure*}[!htb]
\centering
\includegraphics[width=1.0\columnwidth]{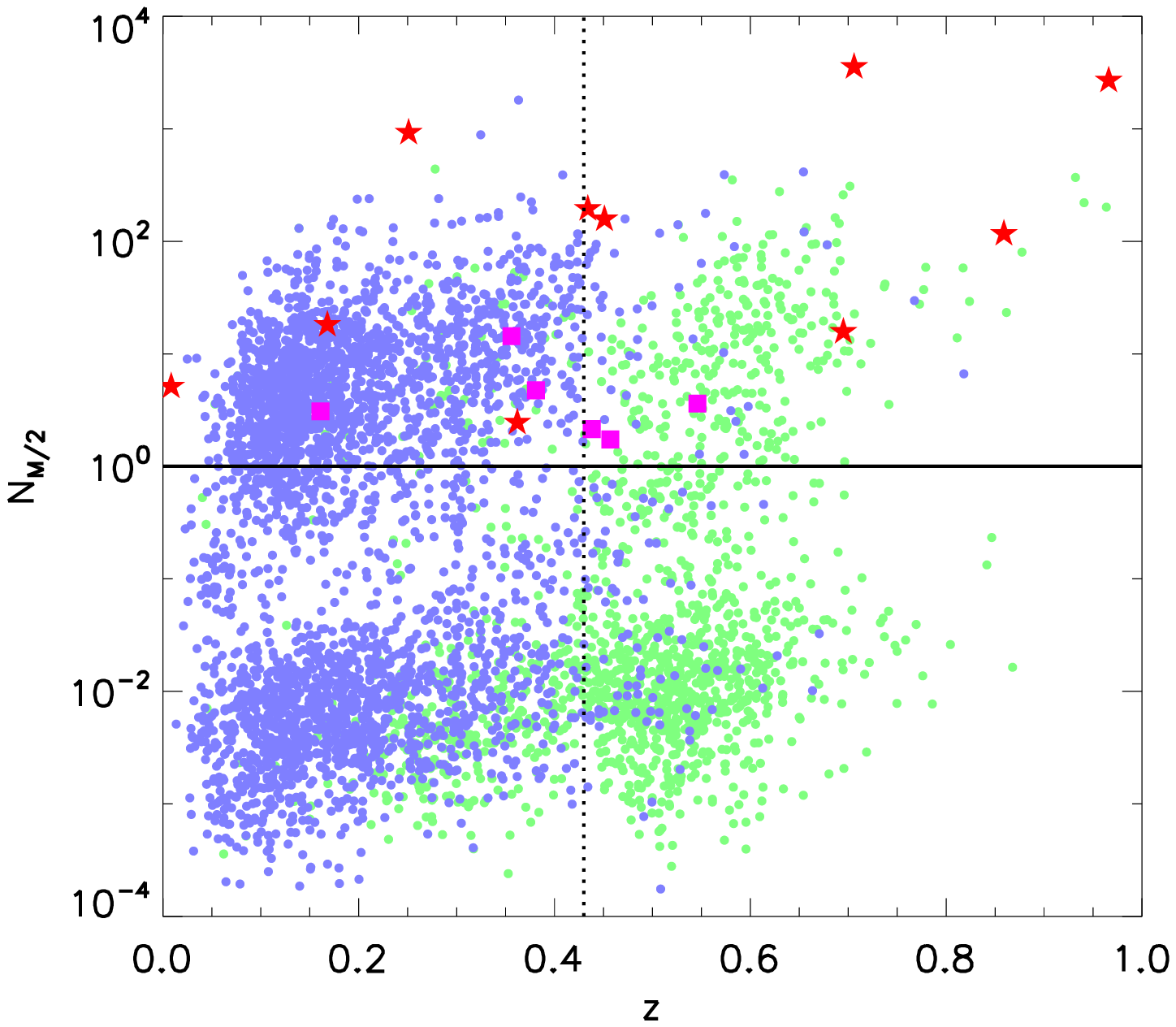}
\includegraphics[width=1.0\columnwidth]{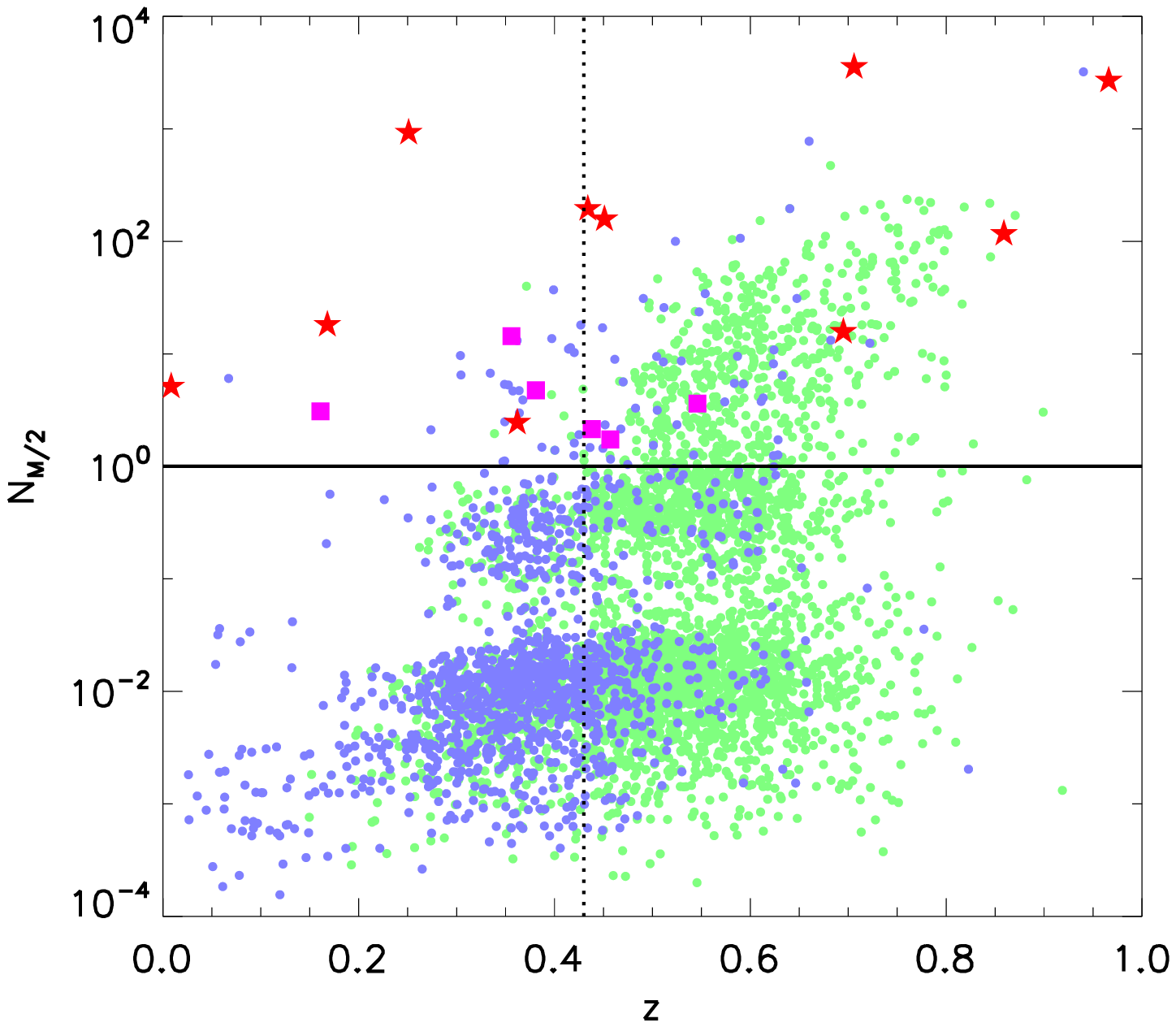}
\caption{
The number of lethal GRBs within 500 Myr at the half mass radius $N_{\rm M/2}$ 
in the galaxies at different redshifts. 
Blue points indicate the SDSS DR8 galaxies, 
and green dots indicate the SDSS/BOSS DR12 galaxies. 
Red stars are LGRB host galaxies and magenta squares are SGRB host galaxies.
Left panel: $10^{10}\ M_{\sun} < M_* <10^{11}\ M_{\sun}$;
Right panel: $10^{11}\ M_{\sun} < M_* < 10^{12}\ M_{\sun}$.
GRB hosts are plotted in both panels regardless of their masses.}
\label{fig:zN}
\end{figure*}

\begin{figure*}[!htb]
\centering
\includegraphics[width=1.0\columnwidth]{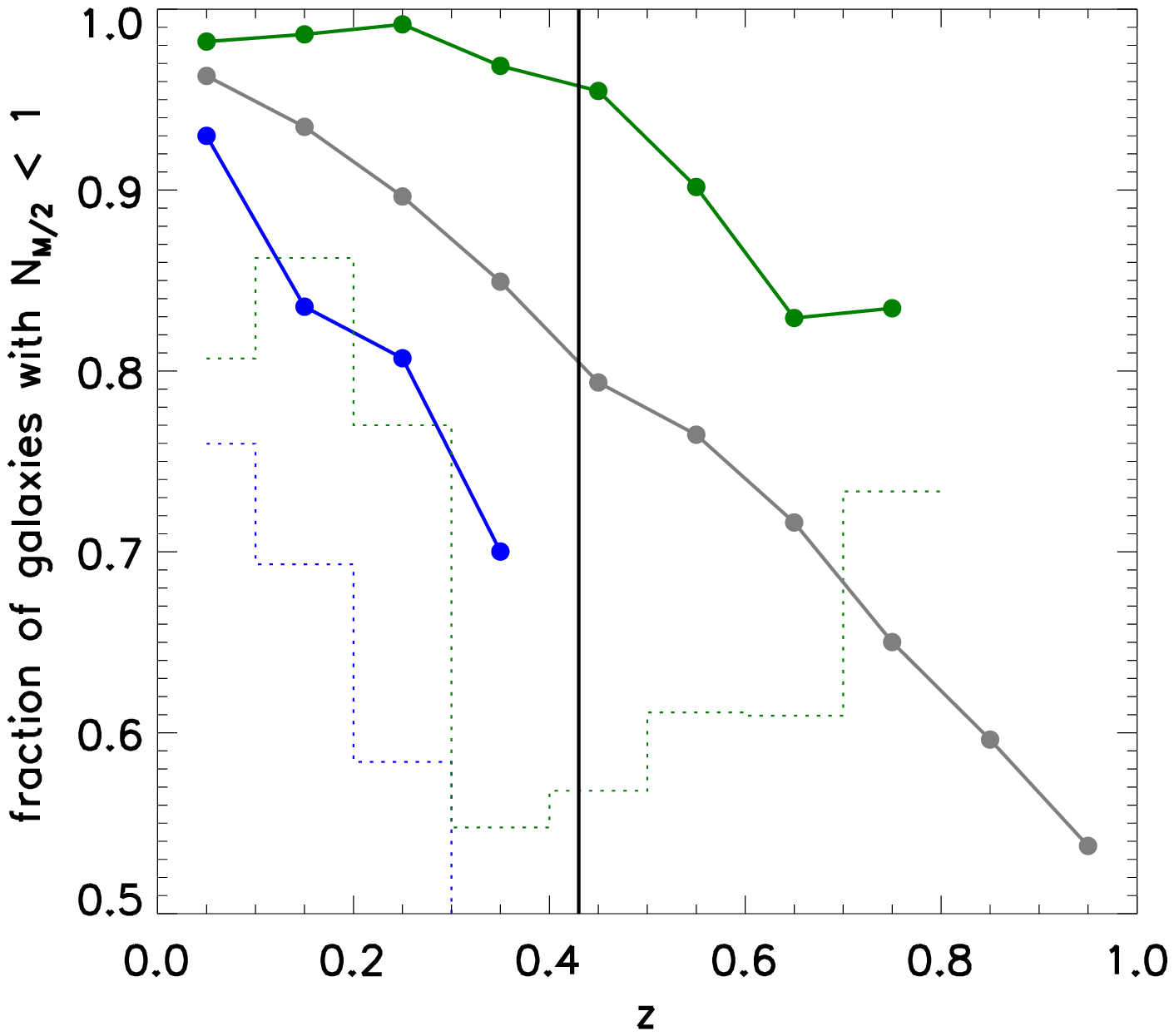}
\includegraphics[width=1.0\columnwidth]{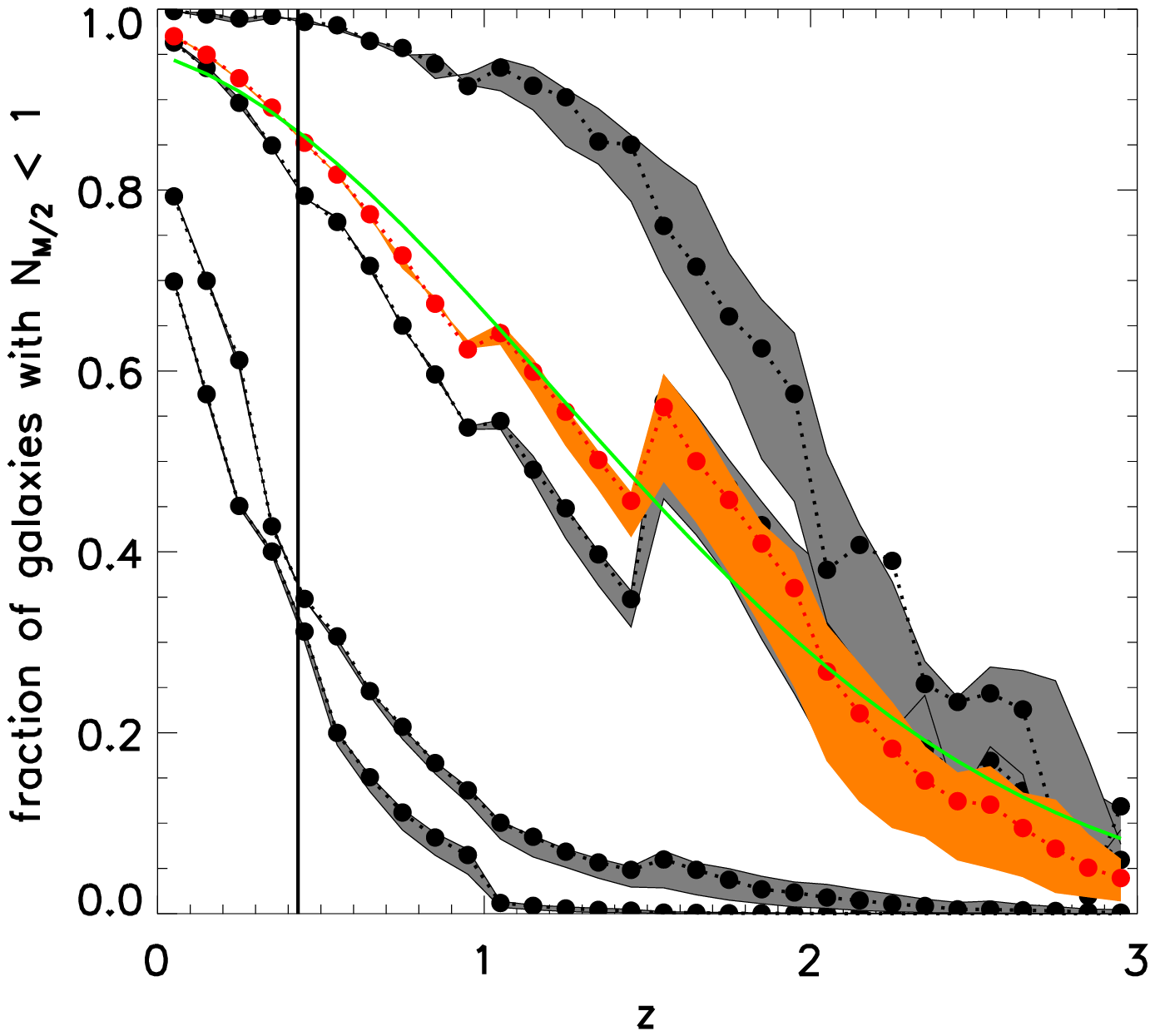}
\caption{
Left panel: The fraction of benign galaxies with $N_{\rm M/2} < 1$ as a function of redshift for the galaxies. SDSS DR8 galaxies (blue line) and SDSS/BOSS DR12 galaxies with $10^{10} < M < 10^{11}\ \rm M_{\sun}$ (green line) are shown. Monte Carlo simulated galaxies in the same mass bin (grey) are also shown for comparison.
Right panel: The fraction of benign galaxies as a function of redshift according to Monte Carlo simulations.
Different black lines denote different mass bins:
from bottom up: $10^{8} < M_* < 10^{9}\ M_{\sun}$,
$10^{9} < M_* < 10^{10}\ M_{\sun}$, $10^{10} < M_* < 10^{11}\ M_{\sun}$
and $10^{11} < M_* < 10^{12}\ M_{\sun}$, respectively.
Red line shows the mass weighted fraction, which is similar to the result of mass bin $10^{10} < M_* < 10^{11}\ M_{\sun}$. The shaded regions denote the uncertainty range of the sSFR factor between red and blue galaxies for each redshift bin (grey) or for mass-weighted case (orange).
The green line shows a Gaussian fit with $G({\rm peak=0.96}, \mu=-0.24, \sigma=1.44)$.
}
\label{fig:frac}
\end{figure*}

The SDSS DR8 sample includes all spectroscopically
classified galaxies in SDSS-I/II. 
It covers 9274 $\rm deg^2$ of the sky and includes 860,836 unique galaxies,
with a median redshift 0.12.
The SDSS/BOSS DR12 sample is the SDSS-III's Baryon Oscillation Spectroscopic Survey
(BOSS) galaxy catalog, which includes a LOWZ sample and a CMASS sample.
There are 1,376,823 galaxies in total, with a median redshift 0.5.

These samples are subject to observational selection effects. 
Since our primary aim is to check whether life can
survive GRBs at high redshifts, for the first step we work on the observed
sample only to make the case.

We use the SDSS DR8 and SDSS/BOSS DR12 galaxy properties obtained from the Portsmouth 
Group\footnote{http://www.sdss.org/dr12/spectro/galaxy\_portsmouth/}.
In this sample, stellar mass, star formation rate, and metallicity 
are estimated by fitting the $ugriz$ photometric data 
with the publicly available code Hyper-Z
\citep{2000A&A...363..476B}.
The stellar population models of
\cite{2005MNRAS.362..799M} and \cite{2009MNRAS.394L.107M} are employed.
Star formation rate is estimated with {an}
exponentially declining star formation history (SFH), 
truncated SFH, or constant SFH {model}, 
whichever gives the best fit. The Initial Mass Function is Kroupa
\citep{2001MNRAS.322..231K}.

Figure \ref{fig:lifemap} displays two 3-D maps of the galaxies in the samples.
For clear illustration only galaxies with {declinations} in the range of
[$0^{\circ}$, $0.5^{\circ}$] are shown here. 
The left panel shows the galaxies with stellar mass 
$10^{10}\ M_{\sun} < M_* < 10^{11}\ M_{\sun}$,
and the right panel shows the galaxies with
$10^{11}\ M_{\sun} < M_* < 10^{12}\ M_{\sun}$.
DR8 galaxies are shown as filled circles and 
DR12 galaxies are shown as open circles.
Colors are encoded with the estimated number of lethal LGRBs
at half mass radius $N_{\rm M/2}$ in 500 Myr.
``Blue'' and ``green'' galaxies have $N_{\rm M/2} < 1$, so that they are safe for life. 
``Red'' and ``orange'' galaxies have $N_{\rm M/2} > 1$, which are {the} most dangerous. 
``Yellow'' galaxies are MW-like galaxies.
One can see that although ``red'' and ``orange'' galaxies gradually increase with
redshift, there are still ``blue''/``green'' galaxies up to redshift 0.7.
The naive speculation that no life can survive beyond $z=0.5$ is not confirmed.

In order to take a closer look at how galaxies become more dangerous at higher redshifts,
in Fig.\ref{fig:zN} we display $N_{M/2}$ of galaxies as a function of redshift.
The left and right panels show the galaxies with stellar mass in the range of
$10^{10}\ M_{\sun} < M_* < 10^{11}\ M_{\sun}$,
and $10^{11}\ M_{\sun} < M_* < 10^{12}\ M_{\sun}$, respectively.
In each panel, DR8 galaxies are shown as blue points and 
DR12 galaxies are shown as green points. 
The vertical dashed line shows the corresponding redshift ($z=0.43$) 
when Earth was formed.
The horizontal solid line marks $N_{\rm M/2}=1$, 
above which most regions in the galaxy are dangerous for life.
One can see in general for both mass ranges, galaxies become more
dangerous at higher redshifts. There seems to be an offset (especially in the 
left panel) between the DR8 (blue) and DR12 (green) galaxies, but it is caused by
the different selection effects for the two samples.

The fraction of galaxies with $N_{\rm M/2} < 1$ is shown 
in the left panel of Fig. \ref{fig:frac}.
The blue and green lines are for DR8 and DR12, respectively.
In order to avoid the selection effects on luminosity and mass,  
we only focus on one mass bin
$10^{10} M_{\sun} < M_* < 10^{11} M_{\sun}$ here.
The results have similar trends for other mass bins.
Each redshift bin is required to have more than 5,000 galaxies.
Some galaxies are marked as SFR = 0. Most of them have a truncated SFH.
Some others have a large age for an exponential decay of SFH as a 
function of time.
For each redshift bin, the fraction of galaxies with SFR = 0
are plotted as the dotted histograms.

Since the DR8 sample tends to have more blue (late type) galaxies for a particular 
stellar mass bin whereas the
DR12 sample originally searched for red (early type) galaxies, the two samples
are both subject to a sample selection effect. In any case, the fraction
lines (blue and green) derived from these two samples 
set the lower and upper limits of
the fraction of benign galaxies as a function of redshift.

\subsection{Monte Carlo Simulations}

In order to reduce the sample selection effects and
investigate redshift-dependence of the fraction of benign galaxies 
for life, 
we apply a Monte Carlo simulation {aimed} at achieving a more complete and 
unbiased sample.

We simulate blue and red galaxies
separately since they follow different correlations
among redshift, stellar mass, sSFR and metallicity
\citep{2009MNRAS.394.1131K, 2015MNRAS.447....2M}. 
We simulate 100,000 blue galaxies in each $\Delta z=0.1$ redshift bin.
The number of red galaxies is estimated by the ratio between {the} number of red and blue galaxies.
The ratio is estimated by integrating the stellar mass function 
of blue and red galaxies from $10^8 - 10^{13}\ M_{\sun}$
in each redshift bin
\citep{2015MNRAS.447....2M, 2013ApJ...777...18M}, which gives
an empirical relation log$_{10}(r/b)=-0.16-0.74z$.

For blue galaxies in each redshift bin,
galaxy mass is given by stellar mass function 
as a function of redshift
\citep{2015MNRAS.447....2M}.
The sSFR distribution is considered as a Gaussian distribution.
The median sSFR is a function of redshift and stellar mass
(Eq.5 of \citealt{2014arXiv1410.4875I}).
The dispersion $\sigma_{\rm sSFR}$ is given as 0.4 Gyr$^{-1}$, a mean value of their Fig. 6.
The metallicity distribution is also assumed to be a Gaussian function.
The median value is a function of stellar mass and SFR (Eq.4 of \citealt{2010MNRAS.408.2115M}).
In order to avoid the abnormal increase of very small and very large $\mu_{0.32}$,
the fourth-order {term} of that equation is dropped.
A dispersion $\sigma_{\rm Fe}=0.07$ is used, which is the median value of the
dispersions in those stellar mass-SFR bins.

The mass distribution of red galaxies as a function of redshift in
\cite{2015MNRAS.447....2M}
is used to simulate the masses of the red galaxies.
They are mostly passive galaxies \citep{2001AJ....122.1861S}
and their sSFRs are generally 0.01 to 0.001 times {of} those of blue galaxies
\citep{2009MNRAS.394.1131K}.
One may assume that their sSFRs follow the same relation with stellar mass
\citep{2009MNRAS.394.1131K}, but
have sSFRs a factor 0.001/0.003/0.010 {of} those of blue galaxies with 
the same masses. The effect of different sSFRs of red galaxies 
are shown as the shadows in the right panel of Fig.\ref{fig:frac}.

The fraction of MC simulated galaxies with $N_{M/2} < 1$ is shown in 
the right panel of Fig.\ref{fig:frac}. 
Black lines are for different stellar mass bins.
From {the} bottom up, the lines are for $10^{8} - 10^{9}\ M_{\sun}$, 
$10^{9} - 10^{10}\ M_{\sun}$, $10^{10} - 10^{11}\ M_{\sun}$
and $10^{11} - 10^{12}\ M_{\sun}$, respectively.
The shaded regions enclose the results for different median sSFRs ratio
between red and blue galaxies. The center line is for 0.003,
while the lower and upper ones are for 0.010 and 0.001, respectively.
It reveals that the fraction of benign galaxies in the
small galaxy sample ($M_* < 10^{10}\ M_{\sun}$) is low, which
becomes smaller than 50\% at $z=0.5$. The fraction of benign galaxies
increases with mass bins. For example, for $M_* > 10^{10}\ M_{\sun}$,
the fraction is still 50\% at $z \leq 1.5$.
The red line and the orange shaded region show the mass-weighted fraction of 
benign galaxies with $N_{\rm M/2} < 1$.
They are similar to the fraction line and shaded region for MW-like galaxies in the mass bin
$10^{10}\ M_{\sun} < M_* < 10^{11}\ M_{\sun}$. This is
because these galaxies are massive and relatively common, and
occupy most of the mass in each redshift bin. 
The fraction of benign galaxies 
is $\sim 50\%$ at $z \sim 1.5$, and is $\sim 10\%$ even at $z \sim 3$.

\subsection{GRB host galaxies}
LGRBs tend to reside in late type galaxies with high star formation rate
and low metallicity \citep[e.g.][]{2009ApJ...691..182S}, whereas
SGRB host galaxies tend to be more diverse \citep{2013ApJ...776...18F, 2014ARA&A..52...43B}.
We study the GRB host galaxies, both for LGRBs and SGRBs, for their {habitability}. Table \ref{tb:grbhosts} lists the information of some LGRB and SGRB
host galaxies with the desired information, including
stellar mass, SFR, metallicity, and scale length.
Also listed are the $N_{\rm M/2}$ values of those galaxies.
They are also overplotted as red stars (LGRB host galaxies) 
and magenta squares (SGRB host galaxies) in Fig.\ref{fig:zN}.
One can see that LGRB host galaxies are among the extremely dangerous
galaxies for life. The {number} of lethal LGRBs at half mass radius within 
500 Myr in all of the LGRB host galaxies {is} much greater than 1, suggesting
that life can hardly survive in these galaxies.
Meanwhile, SGRB hosts are relatively safer than LGRB hosts
due to their much higher metallicity and lower sSFR.
However, in all cases studied in our sample 
one has $N_{\rm M/2} > 1$, so that they still belong to
dangerous galaxies.

\begin{deluxetable*}{llllllll}\label{hosts}
\tablecolumns{8} \tablewidth{0pc} \tabletypesize{\scriptsize}
\tablecaption{GRB host galaxy properties} 
\tablehead{ 
\colhead{GRB} &\colhead{$z$} & \colhead{${\rm log}M$} & 
\colhead{sSFR} &\colhead{12+} & \colhead{$r_{\rm d}$} &
\colhead{$N_{\rm M/2}$} & \colhead{Ref}\\
\colhead{} & \colhead{} & \colhead{$M_{\sun}$} & 
\colhead{ $\rm Gyr^{-1}$} & \colhead{log(O/H)} & \colhead{kpc} &
\colhead{} & \colhead{}
} 
\startdata

    970228& 0.695&  8.65&   1.19&  8.47&  2.458&    15.74& 1,2,3\\
    980425& 0.009&  8.68&   0.54&  8.16&  3.266&     5.14& 1,2,3\\
    980703& 0.966& 10.00&   1.66&  8.14&  1.342&  2696.83& 1,2,3\\
    990712& 0.434&  9.29&   1.23&  8.10&  1.591&   194.86& 1,2,3\\
    991208& 0.706&  8.53&  13.34&  8.02&  0.344&  3558.73& 1,2,3\\
    010921& 0.451&  9.69&   0.51&  8.15&  1.903&   157.67& 2,3\\
    11121 & 0.362&  9.81&   0.35&  8.60& 12.671&     2.46& 2,3\\
    020903& 0.251&  8.87&   3.57&  8.22&  0.431&   927.26& 2,3\\
    030329& 0.168&  7.74&   2.00&  7.97&  0.804&    18.18& 2,3\\
    040924& 0.859&  9.20&   1.19&  8.23&  2.382&   117.08& 2,3\\\hline

        SGRB\\\hline
   050709 & 0.161&  8.80&   0.24&  8.50&  2.080&    3.08& 4,5\\
   051221A& 0.546&  9.40&   0.38&  8.80&  2.290&    3.62& 4,5\\
   061006 & 0.438&  9.00&   0.24&  8.60&  3.220&    2.14& 4,5\\
   070724A& 0.457& 10.10&   0.20&  8.90&  3.640&    1.73& 4,5\\
   071227 & 0.381& 10.40&   0.02&  8.50&  4.720&    4.75& 4,5\\
   130603B& 0.356&  9.70&   0.34&  8.70&  2.020&   14.36& 4,5,6

\enddata
\tablerefs{ 
(1)\citet{2002AJ....123.1111B}; 
(2)\citet{2007ApJ...657..367W}; 
(3)\citet{2009ApJ...691..182S}; 
(4)\citet{2010ApJ...725.1202L}; 
(5)\citet{2013ApJ...776...18F};
(6)\citet{2014A&A...563A..62D}
}
\label{tb:grbhosts}
\end{deluxetable*}

\section{Conclusions and Discussion}

In this paper, we examined the duty cycle of lethal GRBs 
($F_c = 10^8\ \rm erg\ cm^{-2}$) in {the} Milky Way
as well as observed galaxies at different redshifts.
The duty cycle of lethal GRBs attacking Earth is about 1 per 500 Myr,
consistent with the time scale of {the Ordovician} mass extinction. Our existence
suggests that such a duty cycle is long enough to allow advanced life
(such as human beings) to survive, so that such a duty cycle can be regarded
as a conservative criterion for a benign environment. Adopting such a criterion,
we investigated the fraction of benign galaxies as a function of {redshift}
using the SDSS DR8 and SDSS/BOSS DR12 samples and through MC simulations.
We find that this fraction is as high as 99\% in the local universe,
suggesting that the current era is most suitable to {the development of} advanced life.
As expected, the benign galaxy fraction decreases with increasing redshift
due to the increase of SFR and decrease of metallicity at progressively higher
redshifts. However, contrary to the naive expectation that no life can survive
beyond $z>0.5$, we show that benign galaxies do
exist at redshifts much higher than 0.5. In particular, the fraction is 
$\sim 50\%$ around $z \sim 1.5$, and is still $\sim 10\%$ even at $z \sim 3$.
In view that the birth of Earth itself corresponds to $z \sim 0.43$, our results
raise the exciting possibility that advanced civilizations are in principle
not excluded even before the formation of the solar system, even though
the fraction of ``habitable'' galaxies is smaller in the higher redshift universe.

Finally, we discuss some possible uncertainties {inherent} to this analysis. 
{1. Many uncertainties are involved in making the argument that
a GRB is lethal. These include how the atmosphere reacts to intense $\gamma$-ray
flux, how DNA/life reacts to exposure to UV light, and how alien life forms may be different
from the ones of which we are aware. On the other hand, since our analysis focuses on how life
{\em survives} a GRB, these uncertainties do not affect our argument. The only
biological connection used in this paper is our existence, suggesting that the
local lethal LGRB event rate (1 per 500 Myr) can be regarded as a safe (conservative)
duty cycle for life to survive. If the GRB damage is less severe, as suggested by
some recent studies \citep[e.g.][]{2015AsBio..15..207T}, or if advanced life can
tolerate a higher lethal GRB rate (argument presented in the second to last paragraph
in introduction), then the fraction of benigh galaxies at high-$z$ is even larger.
2. There exist uncertainties in the local GRB event rate density, which affect
the estimated lethal GRB duty cycle at Earth. We used the latest results of 
\cite{Sun15} with the largest LGRB sample. 
Adopting the GRB event rate results derived by other authors would 
give a local lethal GRB rate generally consistent with our value, 
with a difference at most a factor of 2-3.}
3. The two galaxy samples included in our study
are subject to substantial selection effects. The derived fractions,
even with MC simulations, may not fully represent the true fractions due to some
un-modeled biases. In any case, we have argued that the DR8 and DR12
samples would bracket the true population. 
The fact that at least some galaxies are benign against
lethal GRBs at high redshifts is robust, which is the main conclusion of this paper.
4. The dependence of LGRBs on SFR and metallicity is evidenced by the 
observations, but the exact correction factors (especially the metallicity one
$f_{\rm Fe}$) are not fully constrained. In any case, the general conclusion of this 
paper does not depend on the concrete form of $f_{\rm sSFR}$ and $f_{\rm Fe}$.
5. We did not study the impact of SGRBs in other galaxies. For most star
formation galaxies, their contribution to lethal GRB number is negligible. However,
for early-type galaxies where SFR is extremely low, SGRBs may dominate the lethal
GRB attacking duty cycle. In any case, $N_{\rm M/2}$ is always much less
than unity in these early-type galaxies. 
For the purpose of this paper (to claim the fraction of benign galaxies at high
redshifts), ignoring the SGRB contribution does not affect the conclusions
and is therefore justified.
{6. We did not study the life-damaging effect of other transients such as
supernovae, which may have comparable damaging effects as GRBs in the local universe
\citep[e.g.][]{2011AsBio..11..343M, 2012MNRAS.423.1234S}. While a more detailed study
is needed to address the supernova effect, the general conclusion of this paper may
not be modified when supernovae are considered. 
This is because supernovae also follow star formation 
history, so that they may be subject to the same $f_{\rm sSFR}$ correction factor.
On the other hand, their metallicity dependence is less apparent than LGRBs, so that
their $f_{\rm Fe}$ correction factor, if any, should be smaller than that of LGRBs.
Overall, their rate of increase with redshift is slower than LGRBs, so that their
role in damaging life at high-$z$ would be less significant than LGRBs.}

\acknowledgments
This work is partially supported by NASA through grants NNX14AF85G and NNX15AK85G.
We thank Houjun Mo, Kentaro Nagamine, Yuu Niino, Tsvi Piran, Liang Qiu, Jared Rice, Hui Sun, and 
Qiang Yuan for helpful discussion, and an anomynous referee for helpful comments.
We also acknowledge the public data available at the SDSS-III web site http://www.sdss3.org/.
Funding for SDSS-III has been provided by the Alfred P. Sloan Foundation, the Participating Institutions, the National Science Foundation, and the U.S. Department of Energy Office of Science.


\bibliographystyle{aa}
\bibliography{refs}

\end{document}